\documentstyle[12pt,aaspp4,epsf]{article}

\def\erg{\hbox{erg}}
\def\keV{\hbox{keV}}

\def\year{\,\hbox{year}}

\def\GHz{\hbox{GHz}}

\def\mJy{\, \hbox{mJy}}
\def\microJy{\, \mu \hbox{Jy}}

\def\cm{\, \hbox{cm}}

\def\degree{\, \hbox{degree}}
\def\tco{t_{co}}
\def\tE{t_{\oplus}}

\def\nuE{\nu_\oplus}
\def\1nE{\nu_{\oplus,1}}
\def\2nE{\nu_{\oplus,2}}
\def\inE{\nu_{\oplus,i}}

\def\nuEm{\nu_{\oplus, m}}
\def\FEm{F_{\nu, \oplus, m}}
\def\FE{F_{\nu, \oplus}}

\def\Lsun{L_\odot}
\def\hf1{\hat{f}_1}
\def\fh2{\hat{f}_2}
\def\bp{Paczy\'{n}ski}
\def\mesz{M\'{e}sz\'{a}ros}
\def\iauc{IAU Circular}
\begin{document}

\title{How to Tell a Jet from a Balloon:
  A Proposed Test for Beaming in Gamma Ray Bursts}

\author{James E. Rhoads}
\affil{Kitt Peak National Observatory, 950 North Cherry Avenue,
Tucson, AZ 85719\altaffilmark{1}}\altaffiltext{1}{Postal address: 
 P.O. Box 26732, Tucson, AZ 85726-6732}
\begin{center}
Electronic mail: jrhoads@noao.edu
\end{center}

\begin{abstract}
If gamma ray bursts are highly collimated, radiating into only a small
fraction of the sky, the energy requirements of each event may be
reduced by several ($\sim 4$--$6$) orders of magnitude, and the event
rate increased correspondingly.  Extreme conditions in gamma ray
bursters lead to highly relativistic motions (bulk Lorentz factors
$\Gamma \ga 100$).  This results in strong forward beaming of the
emitted radiation in the observer's rest frame.  Thus, all information
on gamma ray bursts comes from those ejecta emitted in a narrow cone
(opening angle $\sim 1/\Gamma$) pointing towards the observer.  We are
at present ignorant of whether there are ejecta outside that cone or
not.
 
The recent detection of longer wavelength transients following gamma
ray bursts allows an empirical test of whether gamma ray bursts are
well-collimated jets or spherical fireballs.  The bulk Lorentz factor
of the burst ejecta will decrease with time after the event, as the
ejecta sweep up the surrounding medium.  Thus, radiation from the
ejecta is beamed into an ever increasing solid angle as the burst
remnant evolves.  It follows that if gamma ray bursts are highly
collimated, many more optical and radio transients should be observed
without associated gamma rays than with them.  Published supernova
searches may contain enough data to test the most extreme models of
gamma ray beaming.  We close with a brief discussion of other possible
consequences of beaming, including its effect on the evolution of
burst remnants.

\keywords{Gamma rays: bursts --- X-rays: general --- ultraviolet:
general ---  radio: general --- ISM: jets and outflows}
\end{abstract}

\section{Introduction}
Relativistic expansion is a generic feature of fireball models for
gamma ray bursts.  This can be seen in several ways.  First, the
bursts are luminous (peak $\sim 10^{17} \Lsun$ in gamma rays for
cosmological distances) and vary on millisecond time scales.  The
variability requires them to be small (size $\la \hbox{a few} \times 10^{13}
(\Gamma/100)^2 \cm$ [Woods \& Loeb 1995]).  This in turn guarantees
that they exceed the Eddington limit on luminosity of an object in
equilibrium, because the Eddington mass for $L \sim 10^{17} \Lsun$
comfortably exceeds the mass required to form a black hole in a region
$\la 10^{13} \cm$ across.

The observed gamma ray fluences imply total burst energies of order
$10^{52} \erg$ in the gamma ray range ($20$-$2000 \keV$) (e.g., \bp\ \&
Rhoads 1993), assuming spherical symmetry.  Together with the size
limit, this implies energy densities $\ga 10^{12} \erg/\cm^3$.  This
limit is sufficient to imply a large optical depth to
electron-positron pair creation ($\gamma \gamma \rightarrow e^+ e^-$).
Thus, the energy will not escape directly as electromagnetic
radiation, but will be converted into a relativistic wind of pairs and
baryons.  The observed radiation is presumably generated later by
interactions within such a wind (e.g., \bp\ \& Xu 1994; Rees \& \mesz\
1994) or with an ambient medium (Rees \& \mesz\ 1992; \mesz\ \& Rees
1993).  The nonthermal gamma ray spectra provide additional support
for this scenario (Goodman 1986; \bp\ 1986; Fenimore, Epstein, \& Ho
1993; Woods \& Loeb 1995).


If we are to understand the physical origin of the gamma ray bursts,
we need to know about their distance, their luminosity, their
frequency, and their environments.  The recent detection of a variable
optical counterpart to gamma ray burst 970508 (Bond 1997) and the
detection of interstellar absorption lines at redshift $z=0.835$ in
its spectrum (Metzger et al 1997) clearly shows that at least some
bursts are at cosmological distances, while the observed isotropy of the
burst distribution on the sky (Briggs et al 1996) supports a cosmological
origin for the overwhelming majority of the population.
With the distance scale established, the largest remaining uncertainty
in the burst energy is the solid angle into which the bursts radiate.
This is also the dominant uncertainty in the burst event rate.  If
bursters beam their gamma rays into solid angle $\Omega_\gamma$, the
burst energy scales as $\Omega_\gamma / (4 \pi)$ and the event rate as
$4 \pi / \Omega_\gamma$ relative to the case of isotropic emission.

The maximum plausible degree of collimation for electromagnetic
radiation from a burster is opening angle $\zeta_\gamma \sim
1/\Gamma$, where $\Gamma$ is the bulk Lorentz factor of the radiating
matter.  This is the opening angle into which photons emitted
isotropically in the rest frame of the radiating matter travel in the
observer's frame (cf. Rybicki \& Lightman 1979; Pacholczyk 1970).  It
is of course possible for the burst to be less collimated, if the
ejecta move into a cone of opening angle $\zeta_m > \zeta_\gamma$, but
we still see only those photons emitted by matter in the smaller cone
of opening angle $\zeta_\gamma$.

Collimated jets are remarkably common in astrophysical sources.  We
observe them at small scales (protostars) and large scales (radio
galaxies).  The most widely accepted taxonomy of active galactic
nuclei relies on orientation effects in accretion disk plus jet models
to explain a variety of spectral features.  Some sources have been
observed with relativistic bipolar outflows (e.g., Galactic
micro-quasars [Mirabel and Rodriguez 1994]), and these can show a marked
asymmetry in apparent brightness between the approaching and receding
jets that is well modelled as an effect of relativistic beaming
(Mirabel \& Rodriguez 1994; Bodo \& Ghisellini 1995).
Some possible mechanisms for gamma ray bursts allow naturally for beaming,
through either very strong magnetic fields, accretion disks,
or a combination of the two (e.g., the Blandford \& Znajek [1977]
mechanism).

Lower energy transient sources are expected to follow gamma ray bursts
as the fireball ejecta plow into the surrounding medium 
(\bp\ \& Rhoads 1993; Katz 1994;
\mesz\ \& Rees 1997a; Waxman 1997; Wijers, Rees, \& \mesz\ 1997).
The spectrum at a fixed time is generally
modelled as a broken power law, resulting from synchrotron emission
from a power law distribution of electron energies in a magnetic field
at or reasonably near the equipartition value.  The break in the
spectrum shifts to lower frequencies as the burst remnant ages,
primarily because the bulk Lorentz factor of the ejecta decreases,
reducing the relativistic blue shift of the emitted spectrum.
Generally, the break frequency is expected to decrease as a power law
in time since the burst.  While our imperfect understanding of
relativistic, magnetized shocks leaves large uncertainties in these
models, they are now observationally justified by the observed X-ray,
optical, and radio counterparts to bursts 970508 (IAU circulars
6654--6663) and 970228 (Wijers et al 1997, and references therein).

Because the shift to lower frequencies accompanies the shift to lower
bulk Lorentz factors, the minimum solid angle into which the transient
can radiate increases with time.  This leads directly to our proposed
test for isotropy of gamma ray burst emission.  If bursts are highly
collimated, the gamma rays will radiate into a small solid angle; the
optical transient into a larger one; and the radio transient into a
larger one still.  Thus, we expect to see more optical transients than
gamma ray bursts, and still more radio transients.  On the other hand,
if gamma ray bursts emit isotropically, we do not expect there to be
optical transients unaccompanied by gamma ray bursts.  The ratio of
event rates for burst transients at two frequencies thus gives the
ratio of the mean solid angle into which the burst transients radiate
at those frequencies.  We know the gamma ray burst rate well already,
and within a few months should have a reasonable statistical sample
of optical counterparts to observed gamma ray
bursts.  Establishing the total event rate for all optical transients
with the characteristics of observed burst counterparts (whether 
gamma rays are seen or not) is a large but quite feasible task with
present instruments.

\section{Model-Independent Limits on Beaming}
The simplest model independent form of our test states simply that the
ratio of event rates at two observed frequencies $\1nE$ and $\2nE$ should be
$\hat{N}_1/\hat{N}_2 = \Omega_1/\Omega_2$, where $\Omega_i$ is the
solid angle into which the flux is beamed at frequency $\inE$, and
where $\hat{N}_i$ is the event rate at $\inE$ integrated over all
fluxes.  

Unfortunately, we do not have the luxury of infinitely sensitive
instruments.  We therefore need to replace $\hat{N}_i$ with $N_i$, the
rate of events at $\inE$ exceeding a flux detection threshold
$f_{\hbox{\scriptsize min},i}$.
Observed differences in $N_1$, $N_2$ can then be explained either by
different degrees of collimation at different frequencies, or by
insufficient sensitivity to detect some transients at one or another
frequency.  (The situation becomes more complicated if we are unable
to select a sample of transients caused by a common physical mechanism.)

To account for flux thresholds, consider the joint probability
distribution $p(\hf1,\fh2)$ for a burst to have angle-averaged fluxes
$\hf1$, $\fh2$ at our two frequencies.  (``Angle-averaged'' means that
$\hat{f} = f \Omega / (4 \pi) =  L / (4 \pi d^2)$,
where $f$ is observed flux,
$L$ the source luminosity, and $d$
the luminosity distance to the source.)  Then our observations 
at frequency $\1nE$ will detect a fraction $F_1$ of all transients, where
\begin{equation}
F_1 = { \Omega_1 \over 4 \pi } \times
 \int_{f_{\hbox{\tiny min},1} \Omega_1 / 4\pi}^\infty
 \int_0^\infty p(\hf1,\fh2) d\fh2 d\hf1
~~ .
\end{equation}
A similar equation gives $F_2$, while the fraction of events seen at
both frequencies is
\begin{equation}
F_{12} = { \min{(\Omega_1, \Omega_2)} \over 4 \pi } \times
 \int_{f_{\hbox{\tiny min},1} \Omega_1 / 4\pi}^\infty 
 \int_{f_{\hbox{\tiny min},2} \Omega_2 / 4\pi}^\infty 
p(\hf1,\fh2) d\fh2 d\hf1 ~~.
\end{equation}
We then examine the ratio $F_{12} / F_1$:
\begin{equation}
{ F_{12} \over F_1} = \left[
 \int_{f_{\hbox{\tiny min},1} \Omega_1 / 4\pi}^\infty 
 \int_{f_{\hbox{\tiny min},2} \Omega_2 / 4\pi}^\infty 
 p(\hf1,\fh2) d\fh2 d\hf1
\over
 \int_{f_{\hbox{\tiny min},1} \Omega_1 / 4\pi}^\infty
 \int_0^\infty p(\hf1,\fh2) d\fh2 d\hf1
\right] \times 
{  \min{(\Omega_1, \Omega_2)}  \over  \Omega_1 } ~~ .
\end{equation}
The term in square brackets is $\le 1$ because $p$ is strictly
non-negative.  Also,
$ \min{(\Omega_1, \Omega_2)} / \Omega_1 \le \Omega_2 / \Omega_1$. 
Thus, $\Omega_2 / \Omega_1 \ge F_{12} / F_1$ regardless of the details
of the joint flux distribution $p$.  Likewise, $\Omega_1/\Omega_2 \ge
F_{12} / F_2$.
Although the $F_i$ are defined as fractions of the (unknown) total
transient population, the ratios we care about can be expressed in terms
of measurable event rates: $F_i / F_j = N_i / N_j$.
Combining these results, we find
\begin{equation} 
{N_{12} \over N_2 } \le {\Omega_1 \over \Omega_2 } \le {N_1 \over
N_{12}} ~~.
\label{ineq}
\end{equation}
It is of course necessary that the flux thresholds $f_{\hbox{\tiny
min},1}$, $f_{\hbox{\tiny min},2}$ used to measure $N_{12}$ be the
same ones used to measure $N_1$ and $N_2$.  

If we can constrain $p$ well, we may go beyond this analysis and
actually estimate $\Omega_1 / \Omega_2$.  The resulting estimate will
be sensitive to errors in $p$, however, and such errors will remain
substantial until the lower frequency counterparts of gamma ray bursts are
better studied.  For now, we prefer to emphasize the inequalities
in equation~\ref{ineq}, which are model-independent.  

\section{Expectations in an Illustrative Model}
We now turn our attention to modelling the expected transient event
rate as a function of frequency.  We are most interested in
$\Gamma_p(\nuE)$, the bulk Lorentz factor of the ejecta when the flux
density peaks at observed frequency $\nuE$.  As with most predictions
of relativistic fireball models, $\Gamma_p(\nuE)$ turns out to be a
power law $\Gamma_p \propto \nuE^\mu$ over a large range of
$\nuE$ and $\Gamma_p$.  Published values of $\mu$ cover a substantial
range, from $\mu = 1/4$ (for the ``impulsive fireball'' model in \mesz\
\& Rees 1997a) to $\mu = 9/16$ (from \bp\ \& Rhoads 1993).  
More secure quantitative predictions may be expected later,
when uncertainties in the input physics of the models (particularly
the electron energy spectrum) have been removed by confrontation with
newly available counterpart observations (cf. Waxman 1997).

The detection of radio emission from gamma ray burst 970508 (Frail et
al 1997) at $z > 0.835$ suggests that $\Gamma_p(8.46 \GHz) \ga $ a
few.  Extrapolating boldly to optical and gamma-ray wavelengths, and
using the relatively conservative scaling exponent $\mu = 1/4$, we
infer $\Gamma_p(\hbox{R band}) \ga 30$, and $\Gamma_p(\gamma) =
\Gamma_0 \ga 500$.  This $\Gamma_0$ is consistent with the
observational constraints of Woods \& Loeb (1995).  The maximum
beaming for the gamma ray regime (taking $\Gamma = 500$) is then into
$10^{-6} \times 4 \pi$ steradians.  The optical regime ($\Gamma = 30$)
gives $2.8 \times 10^{-4} \times 4 \pi$ steradians, while the radio
($\Gamma = 2$) gives $0.06 \times 4 \pi$ steradians (though the $\zeta
\propto 1/\Gamma$ scaling is not very accurate at such low $\Gamma$).
The precise values of $\Gamma$ are less important than their ratios.

The observed gamma ray burst rate from BATSE (the Burst and Transient
Source Experiment aboard the Compton Gamma Ray Observatory satellite)
is about 1 burst per day (Meegan et al 1996).
The four bursts well localized by the
BeppoSAX satellite (970111, 970228, 970402, \& 970508) have resulted
in two probable optical counterpart detections.  Thus, the rate of
optical counterparts to observable classical gamma ray bursts is of
order 200 per year, or 1 per square degree per 200 years.  We infer that
if gamma ray bursts are maximally beamed, there should instead be of
order $1$ optical transient per square degree per year, and of order 1
radio transient per square degree per day.

Whether the radio transients predicted under this extreme beaming
scenario would be visible is doubtful: The ratio of their peak radio
flux to gamma ray fluence will be reduced by a factor
$\left[\Gamma_p(\hbox{radio}) / \Gamma_p(\gamma)\right]^2 \sim
10^{-5}$ relative to the isotropic burst case, assuming that all burst
ejecta have about the same initial Lorentz factor.  The reduction of
flux would be less dramatic under scenarios where ejecta have a
wide range of Lorentz factors, like the ``layered jet'' model (\mesz\
\& Rees 1997b; Wijers et al 1997) and ``hypernova'' model (\bp\ 1997).
In these cases, material ejected from the burster at comparatively low
$\Gamma$ might contribute substantially to afterglow flux at low energies
and negligibly at high energies.

\section{Observational Prospects}
The rate of gamma ray bursts is already fairly well known.  We here
consider the requirements for determining the rates of transient
optical and radio events like that observed following gamma ray burst
970508.

The peak optical emission of the 970508 transient was about
 $R = 19.78 \pm 0.05$ magnitude ($37 \pm 2 \microJy$) (Mignoli et al 1997).
The counterpart to burst 970228 was a little fainter ($\approx 17
\microJy$ at V and I bands [Groot et al 1997]).  The events lasted a
few days each.  We will assume that transients of this nature would be
detected by daily observations to limiting $5\sigma$ sensitivity
$R\approx 22$.  Such observations take a few minutes per field with a
1 meter class telescope.  The best field of view presently available
at Kitt Peak National Observatory ($1^\circ$, using the Mosaic CCD camera 
on the 0.9 meter telescope) would allow one to survey roughly 3 fields per
hour, or of order 20 square degrees per night, allowing for overheads.
Thus, event rates
of the order discussed in the extreme beaming scenario could be
tested in $\sim 10$--$20$ nights on the telescope.

Existing optical data may already be sufficient to apply a crude
version of our test.  Deep supernova searches by two groups (the High
Redshift Supernova Search and the Supernova Cosmology Project) are
ongoing and have now detected at least $37$ (Schmidt et al 1996) and
$28$ (Deustua et al 1996; Kim et al 1997) supernovae respectively.
Their search strategy often includes observations separated by
sufficiently short times (a few nights) that optical transients like
the 970508 afterglow would be seen on multiple nights and confirmed as
sources.  We can estimate their total coverage as follows: The
supernova rate is about $34$ per square degree per year for $21.3 < R
< 22.3$ (Pain et al. 1996).  The detected $65$ supernovae therefore
translate to a coverage of about $2 \degree^2 \year$ for supernovae.
The effective coverage for gamma ray burst counterparts may be
somewhat lower owing to the shorter time scale and hence reduced
detection efficiency for these events.  We will take the surveys to
cover $\sim 1 \degree^2 \year$.  The data are therefore sensitive to
optical transient event rates $N_{\hbox{\scriptsize opt}} \ga 1
\degree^{-2} \year^{-1}$. 

Unfortunately, not every transient source detected by these searches
is sufficiently well characterized to say whether or not it could be a
GRB afterglow.  The High Redshift Supernova Search has detected some
short duration transient sources without obvious host galaxies that
might be flare stars in the halo or thick disk of the Galaxy, but
might potentially be GRB afterglows (Schmidt 1997).  Estimating the
expected number of flare star events is possible (see Garibjanian et
al 1990) but beyond the scope of this work.  Observations at a range
of time scales would eliminate this possible source of confusion,
since stellar flares have characteristic durations of minutes to hours
(Krautter 1996), while the afterglows of GRB 970508 and 970228 had
durations of days.

Comparing the previously estimated coverage of these surveys to the
gamma ray burst rate ($N_{\gamma\, \hbox{\scriptsize ray}} \sim 0.01
\degree^{-2} \year^{-1}$), we see that an absence of optical burst
afterglows would imply $\Omega_{\gamma\, \hbox{\scriptsize ray}} /
\Omega_{\hbox{\scriptsize optical}} \ga 0.01$.  This is already
slightly above the ratio $\Omega_{\gamma\, \hbox{\scriptsize ray}} /
\Omega_{\hbox{\scriptsize optical}} \approx 0.004$ predicted in our
illustrative maximal beaming model.  On the other hand, detection of
even one optical transient caused by a gamma ray burster would imply
that $\Omega_{\gamma\, \hbox{\scriptsize ray}} /
\Omega_{\hbox{\scriptsize optical}} \approx 0.01$, i.e., that bursts
are strongly beamed.  In summary, present optical data cover enough
sky for a preliminary application of our test, but firm conclusions
would probably require reanalysis of the data with this test in mind.
More definitive results would be possible with experiments
designed at the outset to search for transients associated with gamma
ray bursters.

In the radio, the FIRST survey is using the Very Large Array to map
the sky at $20 \cm$ wavelengths.  It is achieving limiting
sensitivities of $1 \mJy$ ($\ge 5 \sigma$) in 3 minutes per
observation over a field $\sim 13 \arcmin$ in radius (Becker, White,
\& Helfand 1995).  The observed radio flux of the  970508 transient
has not yet peaked, but seems likely to peak well above $1 \mJy$.
Thus, the event rate in radio can be constrained to about the same
level as the optical rate in a comparable amount of telescope time.
Because the difference in event rates increases as the wavelength
baseline increases in relativistic beaming models, a radio search
could place the most stringent constraints on gamma ray burst beaming.

\section{Discussion}
In addition to its effect on the transient rate at different
wavelengths, beaming of gamma ray bursts may have a few other
observable consequences.  One is that we will sometimes
be near the center line of the jet and sometimes at its edge.  Yi
(1994) pointed out that this will broaden the apparent luminosity
function of the bursts substantially, and explored the implications of
this effect for the statistical properties of the gamma ray burst
population.

Another consequence is for the light curve of any individual burst.
The predicted strength of low energy transients is greatly reduced in
extreme beaming scenarios.  If the burst is beamed into an angle
$\zeta_m > 1/\Gamma_0$, we can expect a qualitative change in the
behavior of the transient when the bulk Lorentz factor drops to
$\Gamma < 1/\zeta_m$.  Before this time, the burst will obey the
predictions of an isotropic model, while afterwards, a correction
factor $\sim (\Gamma \zeta_m)^2$ must be applied to all the flux
predictions for the isotropic case.

Moreover, the dynamical evolution of a beamed burst can change
qualitatively once $\Gamma < 1/\zeta_m$.  In a spherical model, or
while $\Gamma > 1/\zeta_m$, the surface area of the expanding blast
wave scales as $r^2$.  However, the burst ejecta and swept-up matter
expand at the sound speed $c_s \sim c/\sqrt{3}$ in the comoving
frame. The transverse size of the ejecta will thus be the larger of $r
\zeta_m$ and $c_s \tco$, where $\tco$ is the time since the burst in a
frame comoving with the ejecta.  The second term dominates for $\Gamma
\ll  1/\zeta_m$, so that the surface area increases more rapidly with
$r$ and the ejecta decelerate more abruptly.  Specifically, 
we find that the bulk Lorentz factor $\Gamma \propto
\exp(-r/r_{_\Gamma})$  for $1/\zeta_m \gg \Gamma \ga 2$, where
$r_{_\Gamma} = \left[ M_0 \Gamma_0 c^2 / (\pi \rho c_s^2)
\right]^{1/3}$.  (Here $M_0$ and $\Gamma_0$ are the initial mass and
Lorentz factor of the ejecta, and $\rho$ is the mass density of the
ambient medium. Note the implicit scaling $M_0 \propto \zeta_m^2$ for
fixed observable gamma ray properties.)  The observer frame time $\tE$
also evolves exponentially with $r$ in this regime, so that most
directly observable properties follow power law evolution, with a
break when $\Gamma \sim 1/\zeta_m$.  We explore the consequences in a
generalization of the \bp\ \& Rhoads (1993) afterglow model, whose
sole spectral feature is a break where synchrotron self-absorption
becomes important.  We concentrate on the regime where the ratio $f$
of swept up mass to initial mass lies between $1/\Gamma_0$ and
$\Gamma_0$.  Under these conditions, the observed frequency of peak
emission $\nuEm$, flux density at that frequency $\FEm$, and apparent
angular size of the afterglow $\theta$ scale as $\nuEm \sim
\tE^{-2/3}$, $\FEm \sim
\tE^{-5/12}$, and $\theta \sim \tE^{5/8}$ before the break, and as
$\nuEm \sim \tE^{-1}$, $\FEm \sim \tE^{-3/2}$, and $\theta \sim
\tE^{1/2}$ afterwards.  Numerical integrations confirm these relations,
though the transition between the two regimes is quite gradual for
$\nuEm$.  Combining these scalings with the spectral shape ($\FE \propto
\nuE^{5/2}$ at $\nuE < \nuEm$ and $\FE \propto \nuE^{-1/2}$ at $\nuE >
\nuEm$) yields predictions for the light curve at fixed observed
frequency.  The most dramatic feature is in the light curve shape for
$\nuE > \nuEm$, which changes from $\FE \sim \tE^{-3/4}$ to $\FE \sim
\tE^{-2}$.  These exponents are generally sensitive to the assumed
electron energy distribution in the blast wave, and so should be taken
as illustrative of the type of break expected in a beaming scenario.
They could be refined by considering models where the electron
population is fit to observed data.

Note finally that the scaling exponents are also sensitive to the
burster's environment.  As an example, the apparent size in the
$\Gamma > 1/\zeta_m$ regime becomes $\theta \sim \tE^{(5-k)/(8-2k)}$
if we take the medium around the burster to have radial density
profile $\rho \propto r^{-k}$.  This implies that high resolution
studies of these sources (e.g., with very long baseline
interferometry) would have to follow the evolution of the source over
a factor $\ga 4$ in angular size (and $\ga 10$ in $\tE$) if they are
to distinguish between expansion into a uniform density medium ($k=0$)
and a wind environment ($k=2$).

To summarize, we propose a new test for beaming of gamma ray bursts.
Observational constraints on beaming will help eliminate the dominant
remaining uncertainty in the total event rate and gamma ray luminosity
of the bursts.  This in turn will help us determine which classes of
energetic events have the correct frequency to cause the gamma ray
bursts.  The observations required are feasible with present
instruments, and existing data may already be sufficient to test the
most extreme beaming scenarios.

\acknowledgments
I wish to thank Bohdan Paczy\'{n}ski, Dave van Buren, Sangeeta
Malhotra, Brian Schmidt, Jeremy Goodman, Greg Kochanski, Dale Frail,
and an anonymous referee for helpful discussions and comments; the
Infrared Processing and Analysis Center for hospitality during the
course of this work; and the BeppoSAX team for building the satellite
that made the search for gamma ray burst counterparts practical.  This
work was supported by a Kitt Peak Postdoctoral Fellowship.  Kitt Peak
National Observatory is part of the National Optical Astronomy
Observatories, operated by the Association of Universities for
Research in Astronomy.

\bigskip

{\footnotesize \noindent
This manuscript has been accepted for publication in {\it The
Astrophysical Journal Letters\/}, 487:L1.  The scientific content of the
published paper will be identical to that of this preprint, although a
few minor stylistic changes introduced in copy editing are not reflected
here.}

\end{document}